\documentclass{webofc}
\usepackage[varg]{txfonts}   
\usepackage{paralist}
\begin{document}
\title{The O$^2$ software framework and GPU usage in ALICE online and offline reconstruction in Run 3}
%
%

\author{\firstname{Giulio} \lastname{Eulisse}\inst{1}\fnsep\thanks{\email{giulio.eulisse@cern.ch}} \and
        \firstname{David} \lastname{Rohr}\inst{1}\fnsep\thanks{\email{david.rohr@cern.ch}} for the ALICE Collaboration 
}

\institute{CERN}

\abstract{%
ALICE has upgraded many of its detectors for LHC Run 3 to operate in continuous readout mode recording Pb--Pb collisions at 50 kHz interaction rate without trigger.
This results in the need to process data in real time at rates 100 times higher than during Run 2. In order to tackle such a challenge we introduced O$^2$, a new computing system and the associated infrastructure. Designed and implemented during the LHC long shutdown 2, O$^2$ is now in production taking care of all the data processing needs of the experiment.
O$^2$ is designed around the message passing paradigm, enabling resilient, parallel data processing for both the synchronous (to LHC beam) and asynchronous data taking and processing phases.
The main purpose of the synchronous online reconstruction is detector calibration and raw data compression. This synchronous processing is dominated by the TPC detector, which produces by far the largest data volume, and TPC reconstruction runs fully on GPUs.
When there is no beam in the LHC, the powerful GPU-equipped online computing farm of ALICE is used for the asynchronous reconstruction, which creates the final reconstructed output for analysis from the compressed raw data.
Since the majority of the compute performance of the online farm is in the GPUs, and since the asynchronous processing is not dominated by the TPC in the way the synchronous processing is, there is an ongoing effort to offload a significant amount of compute load from other detectors to the GPU as well.
}
\maketitle
\section{Introduction}
\label{intro}
\begin{figure}[htb]
\centering
\includegraphics[width=350px]{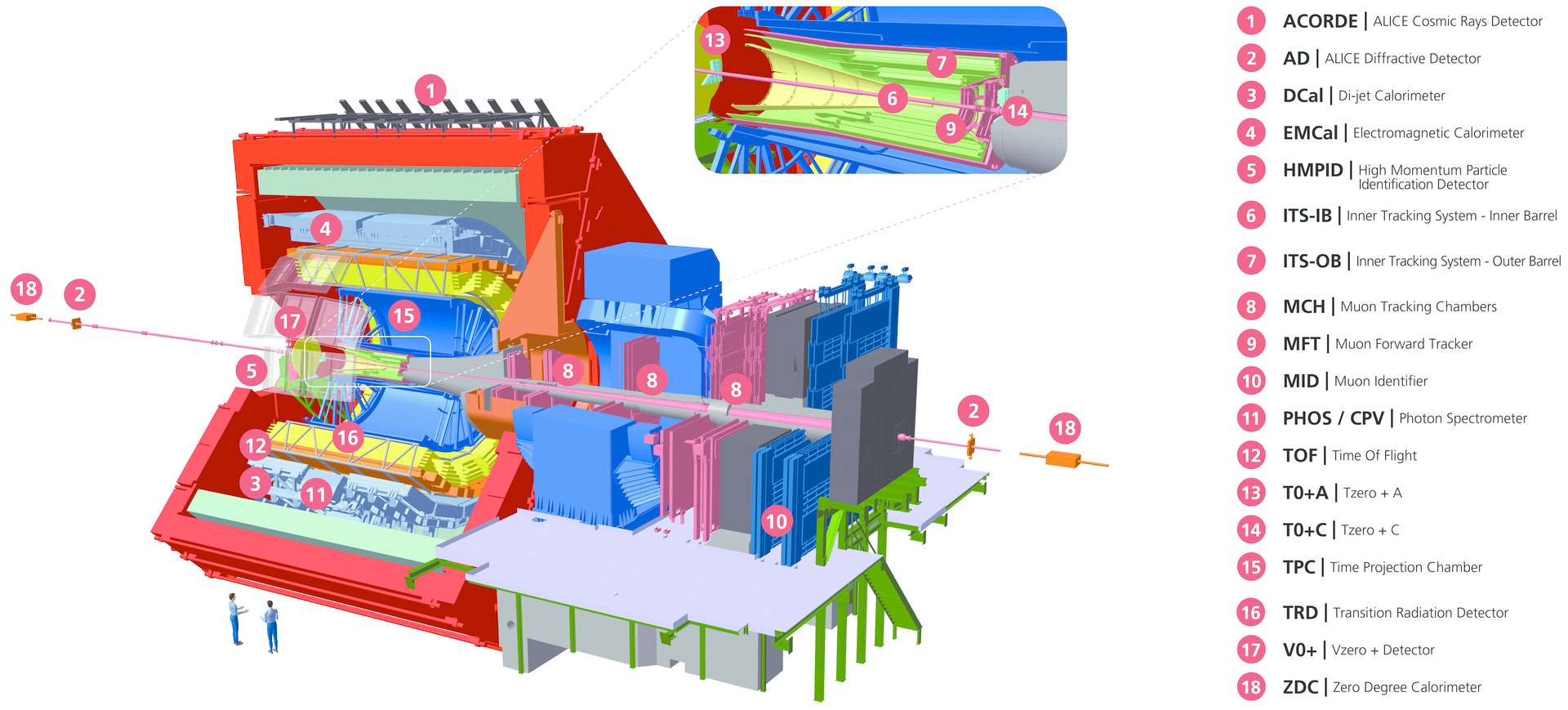}
\caption{Overview of the ALICE Experiment and its detectors in LHC Run 3}
\label{fig:alice}       
\end{figure}

The ALICE experiment at the LHC has been upgraded during the long shutdown 2 (LS2) to operate in continuous readout mode for Pb--Pb collisions at 50 kHz interaction rate without trigger. Fig.~\ref{fig:alice} shows the current configuration of the experiment and lists its subdetectors. This results in the need to process data in real time at rates 100 times higher than during Run 2. In order to tackle this challenge we introduced O$^2$, a new computing system and the associated infrastructure. Designed and implemented during the long shutdown 2, O$^2$ is now in production taking care of all the data processing needs of the experiment. 
In this paper we will concentrate on two aspects of the O$^2$ software stack: the framework itself and the GPU usage in the online and offline reconstruction.

\section{ALICE Run 3 Computing model}
\label{alice-computing-model}

The O$^2$ project has been detailed already in the past \cite{O2TDR}, \cite{DPL}. The system blends the traditional boundaries between online and offline and provides an integrated computing model where the only distinction is whether the input is raw data coming directly from the experiment or compressed raw data read from the storage. In addition, certain processing steps can be enabled or disabled only during synchronous or asynchronous reconstruction, and e.\,g.~different cuts are used to speed up the processing during the synchronous phase. 
During the synchronous phase, data coming from the detector is read out by a first layer of computing nodes, dubbed First Level Processors (FLP), each associated to a subset of the detector readout channels. From this layer the FLP2EPN network assembles the so-called time frames, each containing 2.8 ms of continuous readout data. These time frames are then scheduled to arrive at one of the nodes of a second layer of computing elements, dubbed Event Processing Nodes (EPN). The EPN farm is fitted with 350 nodes, each with 8 GPUs, and performs the synchronous pass of reconstruction on the data.
The goal of the synchronous phase is to reduce the sheer amount of data, in particular by compressing the raw data from the TPC detector, using a combination of lossy and lossless techniques, and to perform the detector calibration. In addition, it does a full quick reconstruction of the data, for some detectors only for a subset of the collisions, which is then used for Quality Control (QC) and Monitoring. The output of the synchronous phase is a set of compressed raw data, denoted Compressed Time Frame (CTF), and the calibration data. For development, debugging, and monitoring reasons, a small subset (less than one per mille) of the raw data is stored in parallel.
CTFs are stored in a 120 PB disk buffer at CERN and eventually replicated over to the Grid.

The data from heavy-ion collisions remains on the disk buffer for roughly one year until the next heavy-ion period.
When there is no beam in the LHC, the data is read back from the CERN buffer to the EPNs for the asynchronous reconstruction.
This phase performs the full event reconstruction of the data using the final calibrations.
In addition to the asynchronous reconstruction on the EPN, WLCG Grid sites also perform the asynchronous reconstruction for a part of the data, irrespective of the LHC status, reading the data from the replicas in the Grid.
The result of the asynchronous reconstruction is a set of Analysis Object Data files (AOD), which are used by the ALICE community to perform the actual physics analysis. It should be noted that the analysis also uses the same software stack. Its description is however outside the scope of this document and more details can be found in previous work of ours~\cite{O2ANALYSIS}.

\section{The O$^2$ framework}

To implement this new computing system, we collaborated with GSI to create a versatile software framework, known as ALFA (ALICE - FAIR). 
This framework features a layered design and it is designed to accommodate our diverse processing requirements.
The foundation is the so-called "Transport Layer" which as the name implies is responsible for moving the data around. It is based on a message passing toolkit named FairMQ \cite{FAIRMQ}.
In FairMQ, the data processing happens in separate entities named devices. Each device is a UNIX process to allow for easy deployment and isolation. Multiple devices can be connected in a so-called topology, where they exchange messages over so-called "channels".

The first advantage of this approach is that it allows for crashing "expendable" devices without the need to kill the whole topology. This is particularly critical in the online environment, where we want to minimise the downtime due to failing optional components, like for certain QC tasks. Moreover the problem gets naturally partitioned and becomes more manageable, e.g. when something crashes we get the core file only from the crashing device.
Although FairMQ is a general purpose message passing toolkit, on the same node data transfers are optimised by using a shared memory buffer for message payloads. In this scenario, devices only exchange references to the data in shared memory, which is crucial for handling the large data payloads coming from the detector.
This optimisation aside, the message passing paradigm allows for seamless integration of devices across networks, therefore providing a unified view for both local and distributed computation.

The second layer is the so-called "Data Layer", which is used to describe the data inside the messages. Multiple backends are supported by the framework, including ROOT \cite{ROOT}, Apache Arrow \cite{ARROW} for data analysis, or detector tailored formats optimised for zero copy and specially targeted for direct usage from the GPU. This allows us to map data directly in the GPU buffers for reconstruction, minimising copies and format changes.
The final layer is the so-called "Data Processing Layer" (DPL). The main role of this layer is to abstract the distributed nature of the system and present a more familiar data flow engine to the end user.

The user describes the data flow in terms of tasks that act on the data, specifying data dependencies and providing the system a hint of what outputs might be produced by a given task. The DPL aggregates all the user-provided task descriptions and produces the FairMQ topology which will implement the data flow.
DPL takes also care of interfacing with the deployment system of the experiment, generating the necessary configuration files to have the workflow deployed on the computing nodes. When a deployment system is not present, like when running asynchronous processing, DPL provides a self hosted deployment system, called the DPL driver, which takes care of starting the tasks in a topology and manages the task lifetimes.
This strategy allows ALICE to have a homogeneous software environment all the way through the processing chains and DPL effectively abstracts away the deployment system from the data processing.

Besides being instrumental to running workflows in batch, the DPL driver also provides an interactive development and debug environment for users to run on their laptop. This mode opens a so-called Debug GUI on the screen when the user runs a workflow. The GUI allows the user to investigate the topology, to inspect each device configuration and logs, and to perform some common manipulations on the selected devices.
For example, it can be used to attach instruments like a debugger or a profiler, to send signals, or to modify the log level of a given device. The GUI also has facilities to quickly visualise the status of the computation and the metrics associated to the processing, providing a quick interface to display them without the need of a full monitoring infrastructure. The GUI is in general run directly on the user's local system, however it has been recently upgraded so that in can work in a remote mode, where a thin HTML client is used to render in a browser window the state of a workflow running on a remote machine.

Internally DPL owns the state machine which is running the workflow, interfacing with the facilities which are provided by FairMQ. In particular, while in the RUN state, DPL has its own event loop which is responsible for handling incoming data requests. The event loop is implemented thanks to OpenSource library \textit{libuv}, originally developed for the NodeJS project and which abstracts the interface with the system socket polling mechanisms, \textit{epoll} on Linux and \textit{kqueue} on macOS / FreeBSD.

\section{Parallelism in the O$^2$ framework}

The actor-model design of FairMQ allows for multiple approaches to parallelism, which are marshalled by DPL. The most natural one is running each device as an independent process, which therefore processes data asynchronously, allowing multiple timeframes to be processed at the same time. This is what we call "horizontal" parallelism. Similarly, independent parts of the processing chain can run at the same time, processing different parts of a given timeframe. This is enabled by the fact that while FairMQ uses a message passing paradigm, when two devices are running on the same machine, the optimised shared memory back-end allows shallow copies of the timeframe data (or parts thereof) without any performance penalty. It will be up to the transport layer to keep track of the reference counting and avoid copies. This allows efficient "vertical" parallelism, i.e. different devices processing the same timeframe in parallel.
Of course, this approach to parallelism, while very natural to implement on top of a message passing toolkit like FairMQ, has the drawback that without feedback data tends to accumulate in the input queue of the slowest device. To prevent this, DPL provides an out-of-band back channel that allows the last device of the chain to notify the first one about its processing status, allowing the latter to implement several rate limiting tactics. 

Moreover, DPL allows a given device to be replicated a number of times. This happens without any extra coding by the user. Data will then be served to each replica in a round-robin manner. This is particularly useful to improve the performance of those steps which are bottlenecks for the data flow, and it resulted in an effective way to improve parallelism in a seamless manner. This has been critical in particular for all devices which are downstream from the one which processes data on the GPU, due to the large data throughput of the latter.

Finally, while DPL at the moment does not provide multithreading support out of the box, the design is already geared to allow multiple algorithms to run in parallel on the same device. We expect the natural evolution of the framework will go in the direction of automatically merging devices running algorithms which would benefit from running together, in a multithreaded fashion. This will however always be considered an optimisation, keeping the original message passing architecture intact.

\section{Computing steps in synchronous and asynchronous reconstruction}

Before discussing how synchronous and asynchronous reconstruction are implemented, it is instructive to get an overview over which processing tasks are running in which phase.
The main purpose of the synchronous processing is compression, calibration, and quality control.

Data compression for all detectors uses a custom rANS~\cite{bib:ranslettrich} library for entropy encoding.
In addition, detector-specific preprocessing steps can reduce the entropy, and some detectors also perform non-lossless first reconstruction steps during synchronous reconstruction to reduce the data volume.
The detector with the largest data volume by far is the TPC.
The TPC chain for data compression consists of zero-suppression after common-mode correction and ion tail cancellation in the FPGAs in the FLPs and cluster finding in the GPUs on the EPNs as non-lossless steps.
This is followed by a track model compression to store cluster to track residuals instead of absolute coordinates~\cite{bib:lhcp2017} and other TPC-specific steps to reduce the entropy.
Finally the TPC data goes through the common ANS compression.
The track model compression requires full tracking of the TPC data for all collisions, thus the synchronous reconstruction must perform the full TPC tracking at sufficient resolution.
Data compression for other detectors also involves clustering, but does not require full event reconstruction.

Each detector of ALICE has several calibration tasks during the synchronous processing, mostly running custom algorithms.
Some of these tasks require reconstructed collisions and tracks, while others are very basic, running directly on the raw data stream.
The prime example for a task that needs tracks, and also the most complicated calibration tasks, is the creation of the correction maps for the TPC space charge distortions~\cite{bib:lhcp2017}.
This requires matching TPC tracks to the inner and outer detectors ITS, TRD, and TOF, and a refit of the tracks without the TPC information to obtain the distortion of TPC clusters with respect to the refitted trajectory.
While this requires full tracking of the barrel region of ALICE, it is sufficient to reconstruct less than 5\% of the most peripheral collisions.
Thus, overall, calibration requires full event reconstruction, but only on a small subset of the collisions of a time frame, and thus the reconstruction required for the calibration is computationally less critical than for the data compression.

QC runs several custom algorithms on the raw data streams of all detectors, while the global quality control is based on fully reconstructed collisions.
Thus, similarly to the calibration, quality control requires full event reconstruction, but only on a small subset of the collisions.

In summary, the synchronous reconstruction requires full tracking of the TPC, full reconstruction and tracking of a small subset of collisions of all detectors, and it runs several detector-specific calibration and quality control tasks.
In contrast, the asynchronous reconstruction performs the full event reconstruction for every collision, it runs similar quality control tasks, but basically no calibration tasks.
Consequently, from a computational perspective, the synchronous processing is fully dominated by TPC tracking, while the computing load during asynchronous reconstruction is distributed much more heterogeneously over all detectors.
Tables~\ref{tab:syncreco} and~\ref{tab:asyncreco} give an overview of the relative processing times, confirming the expectation that synchronous processing is fully dominated by the TPC with a contribution above 99\%.
Note that these numbers were measured using only the CPU version of the algorithms, i.\,e.~without GPU processing, and the multithreading was configured in such a way to have as many time frames in flight as the available memory of the node permits, and then use multi-threading on top to the extent that constant 100\% CPU load is obtained.

\begin{table}[htb]
\centering
\caption{Relative processing times of synchronous reconstruction steps. Calibration and QC tasks are excluded in this table, since they are not relevant for GPU offload.}
\label{tab:syncreco}
\begin{tabular}{lr}
\hline
Processing step & Relative time \\
\hline
TPC Processing (Tracking, Clustering, Compression) & 99.37\% \\
EMCAL Processing & 0.20\% \\
ITS Processing (Clustering + Tracking) & 0.10\% \\
TPC rANS Encoding & 0.10\% \\
ITS--TPC matching & 0.09\% \\
MFT Processing & 0.02\% \\
TOF Processing and Global Matching & 0.02\% \\
\hline
Rest & 0.1\% \\
\hline
\end{tabular}
\end{table}

\begin{table}[htb]
\centering
\caption{Relative processing times of asynchronous reconstruction steps.}
\label{tab:asyncreco}
\begin{tabular}{lr}
\hline
Processing step & Relative time \\
\hline
TPC Processing (Tracking) & 61.41\% \\
ITS--TPC matching & 6.36\% \\
MCH Clusterisation & 6.13\% \\
TPC rANS Decoding & 4.65\% \\
ITS Tracking & 4.16\% \\
TOF Matching & 4.12\% \\
TRD Tracking & 3.95\% \\
MCH Tracking & 2.02\% \\
AOD Production & 0.88\% \\
\hline
Quality Control & 4.00\% \\
\hline
Rest & 2.32\% \\
\hline
\end{tabular}
\end{table}

\section{Design of computing farm and selection of algorithms for GPUs}

The prime design goal of the EPN farm is to sustain the maximum expected data rate during synchronous processing of 50 kHz Pb--Pb data taking.
From the above explanations and Tables~\ref{tab:syncreco} and~\ref{tab:asyncreco} it is clear that the only critical component for synchronous processing is the TPC track reconstruction.
Since ALICE already had good experience with the usage of GPUs in the High Level Trigger (HLT) of LHC Run 2~\cite{bib:hltcpc}, it was decided to run the TPC reconstruction on GPUs, and the bulk of the processing capabilities of the EPN farm should be in the GPUs~\cite{bib:alice2upgrade}.
Placing multiple GPUs in the same server saves costs for the infrastructure, thus ALICE uses 8 AMD MI50 or 8 AMD MI100 GPUs in each server.
Benchmarks with Monte Carlo simulations have shown that 64 (96) physical CPU cores are required to feed the MI50 (M100) GPUs with data. The servers have 512 GB (1 TB) of main memory, and are connected with 100 Gbit InfiniBand.

\begin{figure}[htb]
\centering
\includegraphics[width=350px]{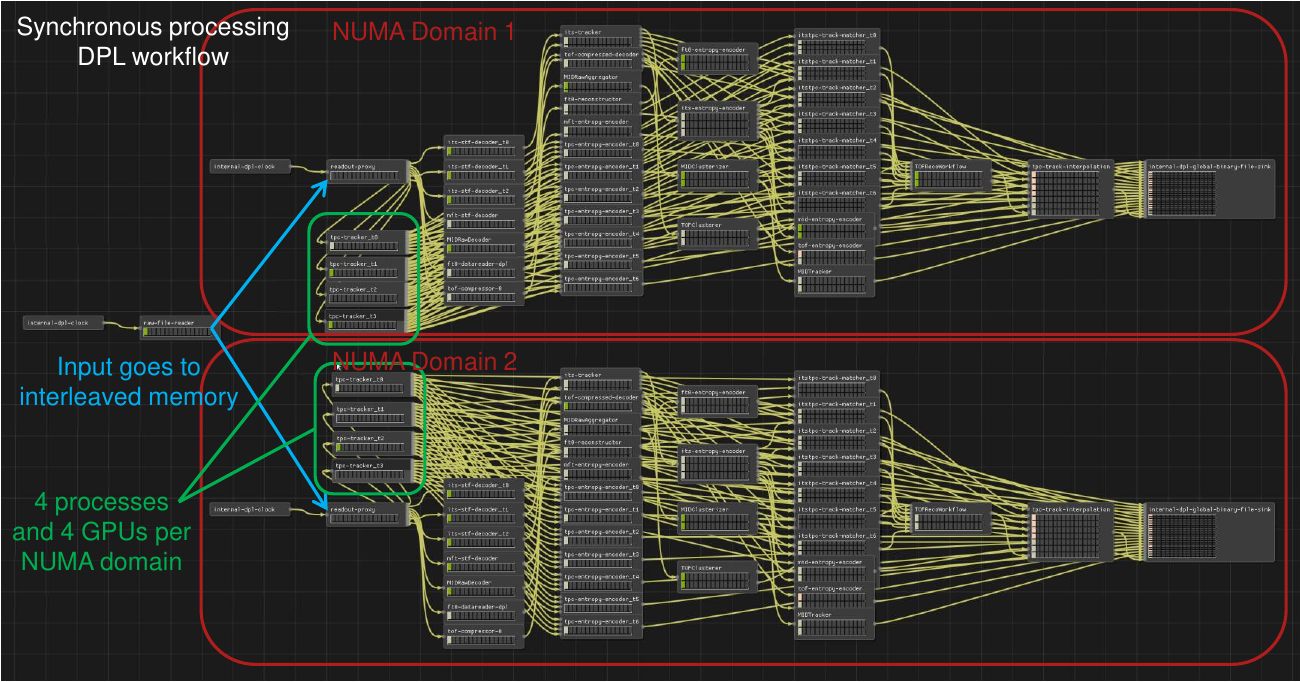}
\caption{Illustration of the ALICE Run 3 synchronous reconstruction workflow for 8 GPUs and 128 virtual CPU cores. The workflow is split in the two NUMA domains, sharing only the shared input buffer. Both NUMA domains have 4 GPUs, which are each driven by individual OS processes. For simplicity, only the reconstruction processes are shown, and QC and calibration processes are omitted.}
\label{fig:workflow}       
\end{figure}

The processing distributes the time frames in a round-robin fashion among the GPUs, avoiding any GPU intercommunication.
In order to also minimise the communication between the CPU sockets of the dual-socket server, the DPL workflows for synchronous and asynchronous processing are split into two workflows, one running per NUMA domain.
The only component that is shared between the NUMA domains is the global input buffer for the raw data, since the server has only one network HCA writing into the global input buffer.
Fig.~\ref{fig:workflow} illustrates the workflow.

\begin{figure}[b]
\centering
\includegraphics[width=350px]{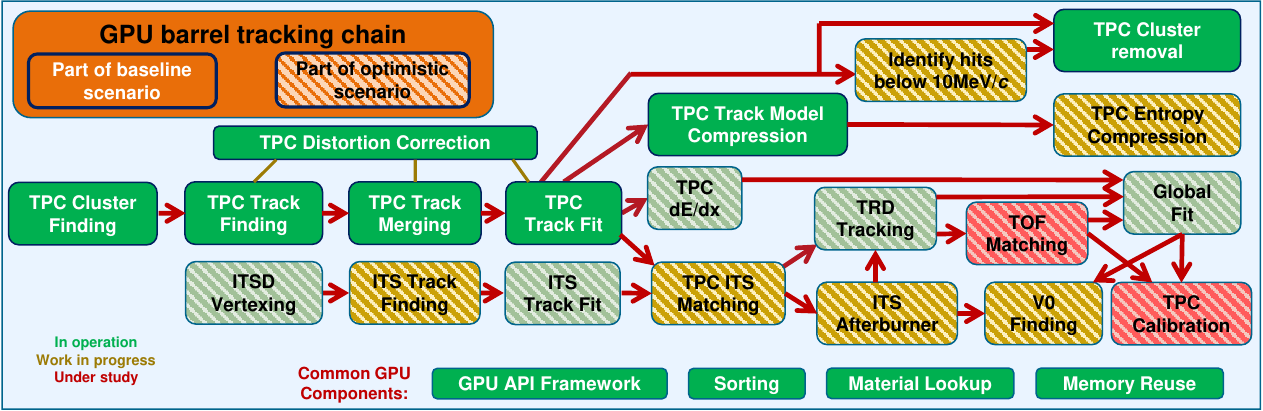}
\caption{Illustration of the processing graph of synchronous and asynchronous GPU processing steps in the baseline and optimistic scenarios. Colors indicate the readiness to run this step on the GPU. All steps are fully implemented and commissioned to run on CPU in the asynchronous processing.}
\label{fig:scenarios}       
\end{figure}

Since the EPN farm will perform asynchronous processing roughly half of the time, good CPU and GPU utilisation in the asynchronous phase are also important aspects.
Thus the GPU software development was following two scenarios:
\begin{compactitem}
 \item The baseline scenario contains all components that are mandatory to run on the GPU for successful Pb--Pb data taking at 50 kHz interaction rate. In addition, steps that were already ready to run on GPU are also included. The baseline scenario had to be implemented, otherwise ALICE could not take data during the heavy-ion run.
 \item The optimistic scenario contains a larger fraction of the steps running on the GPU in order to achieve higher GPU load during the asynchronous reconstruction. It is not necessary for synchronous data taking. It is beneficial to have consecutive processing steps on the GPU to avoid intermediate data transfer back and forth. Thus, the full central barrel global tracking was chosen as candidate for the optimistic scenario.
\end{compactitem}
Fig.~\ref{fig:scenarios} shows the reconstruction steps of the barrel tracking, and which step is considered in which scenario, as well as the ready state of the steps.
While the baseline scenario for synchronous processing is already fully implemented and in operation, ALICE is still working to complete the optimistic scenario to improve the performance in the asynchronous reconstruction.

\begin{figure}[t]
\centering
\includegraphics[width=350px]{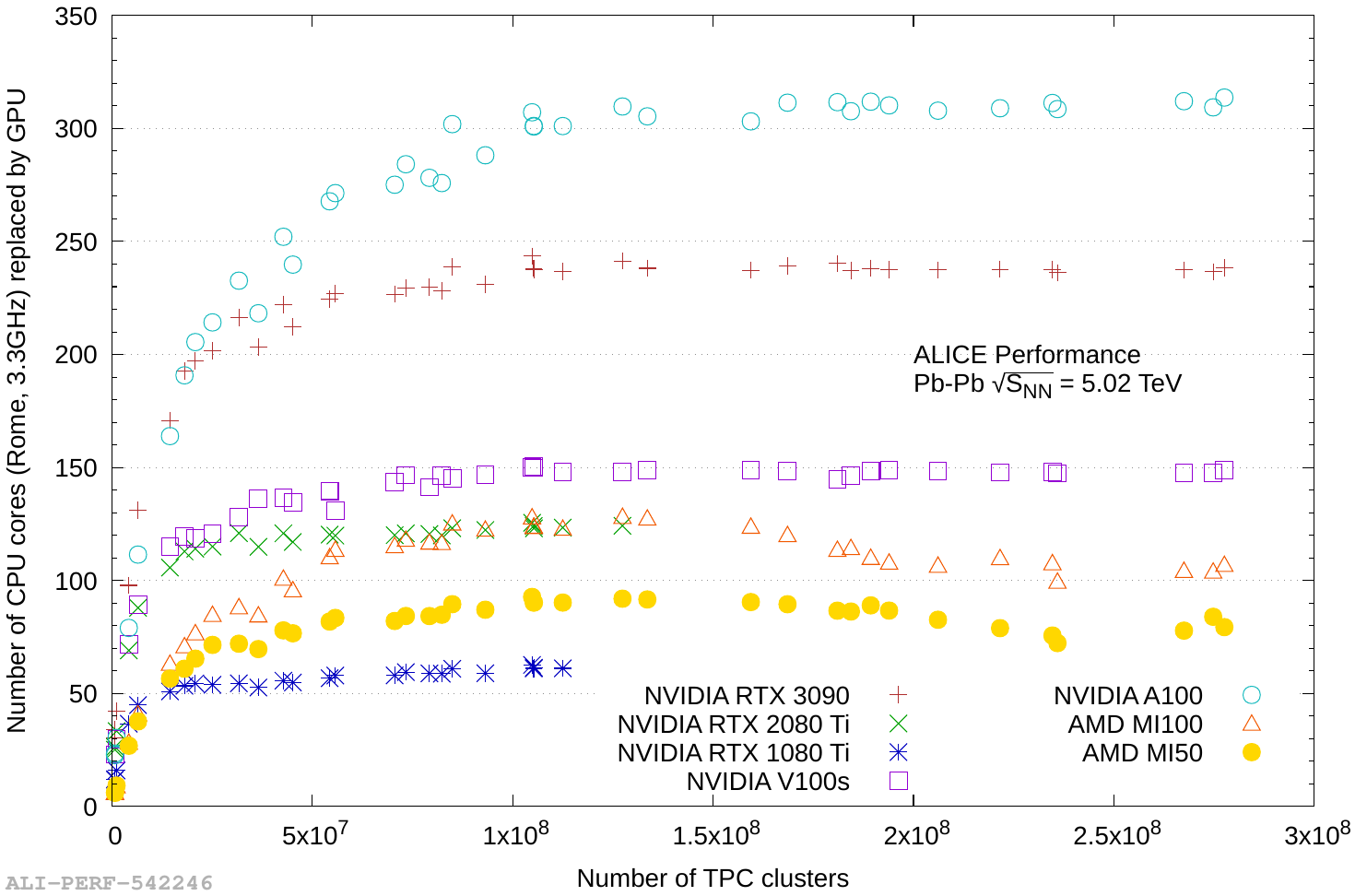}
\caption{Speedup of several GPU models compared to one AMD Rome CPU core in the EPN servers. The measurements are corrected for the number of CPU cores required to drive a GPU, i.\,e.~the figure states how many CPU cores can be replaced by one GPU.}
\label{fig:gpuspeedup}       
\end{figure}

In 2022, it became clear that the original estimates for the number of hits in the TPC with the new detector were lower than in reality.
Since the required computing time goes linearly with the number of TPC hits, an extension of the EPN farm was performed to restore the 30\% computing capacity margin over the 50 kHz Pb--Pb estimates.
While the original EPN servers were equipped with MI50 GPUs as specified above, the new servers have MI100 GPUs and consequently more CPU cores and memory.
Fig.~\ref{fig:gpuspeedup} gives an overview of the GPU models that were considered.

\section{Performance in synchronous and asynchronous reconstruction}

The performance of the TPC tracking for synchronous reconstruction with Pb--Pb multiplicity was extensively benchmarked using Monte Carlo simulations~\cite{bib:chep2021gpu}.
Since then, it turned out that there will be more TPC clusters in the new TPC than predicted and the computing estimates were adjusted accordingly.
Because there was no heavy-ion data taking at high interaction rate in LHC Run 3 yet, MC simulations and extrapolations from the low interaction rate heavy-ion run in 2022 are the only available input.
Further improvements to the physics performance of the tracking slowed down the processing to some extent, which has been mostly compensated by performance improvements of similar size.
Overall, with the latest MC simulation and extrapolations from 2022 low interaction rate data, the current EPN farm with 350 nodes should be able to process the 50 kHz Pb--Pb data with a computing margin of 30\%.
From the performance shown in Fig.~\ref{fig:gpuspeedup} relative to CPUs, a comparable CPU-only computing farm would require more than 2000 servers with 64 physical cores each, which would be prohibitively expensive.

The experience from the pp data taking and the low interaction rate Pb--Pb data taking did not yet reveal severe issues, but also did not stress-test the computing infrastructure to the full extent.
Due to the free available computing resources at low interaction rate and during pp data taking, additional reconstruction steps could be enabled during the synchronous processing for validation and for studies, which will not be possible during Pb--Pb.

The situation for asynchronous reconstruction is quite different.
In the synchronous reconstruction, the EPNs must keep pace with an externally defined input data rate, and keep a certain compute margin for rate fluctuations and unforeseen cases.
In contrast, the asynchronous reconstruction should process the compressed time frames from the disk buffer as fast as possible.
Currently, the asynchronous processing is CPU-bound and the GPU utilisation itself does not play a large role.
Hence, asynchronous reconstruction aims for 100\% CPU utilisation, while the publishing rate of time frames at the beginning of the processing graph is a free parameter.
It should be high enough to achieve full CPU utilisation, but also not too high to avoid over-subscription of the CPU cores and to reduce the memory footprint by having fewer time frames in flight.
The workflow employs a simple rate limiting algorithm, counting how many time frames are in flight globally, and throttles the injection accordingly.
In addition, a smoothing of the publishing rate using heuristic filters is applied, to achieve a more even publishing rate, and avoid waves of high compute load percolating through the processing chain.

In the end, the processing achieves a CPU utilisation of more than 90\%.
Increasing the actual load to a constant 100\% is only possible by having more time frames in flight, which would often oversubscribe the CPU cores, require much more memory, cause more context switches, and actually yield a lower processing throughput.
It should also be noted that the load is measured in virtual hyperthreaded CPU cores, i.\,e.~from 50\% on, all physical CPU cores are loaded, while the second virtual core per physical core provides only around 20\% additional compute capacity.
Thus, the unused compute capacity at 90\% load is not 10\% but around 2\%.

\begin{table}[htb]
\centering
\caption{Processing time per time frame in seconds on an original EPN server with MI50 GPU and 64 physical CPU cores. The second column shows the time per time frame of a single workflow, using the resources listed under configuration. The third column shows the total throughput achievable on the EPN by running multiple workflows in parallel to utilise all resources.}
\label{tab:asyncperf}
\begin{tabular}{lrr}
\hline
Configuration & Time per TF & Time per TF \\
& (1 workflow) & (full server) \\
\hline
8 virtual CPU cores & 76.91s & 4.81s \\
16 virtual CPU cores & 34.18s & 4.27s \\
1 GPU + 16 virtual CPU cores & 14.60s & 1.83s \\
1 NUMA domain (4GPUs + 64 virtual cores) & 3.5s & 1.7s \\
\hline
\end{tabular}
\end{table}

In order to compare CPUs to GPUs, four asynchronous workflow configurations were tuned for the EPNs, two using only CPUs, and two using CPUs and GPUs.
Table~\ref{tab:asyncperf} lists the achieved performance.
The 1 NUMA workflow, running twice per EPN, is the logical setup, resembling the synchronous processing workflow, and turns out to be the fastest configuration as expected.
Comparing e.\,g.~the workflow with one GPU to the 1 NUMA domain workflow, the latter has the advantage that it has to run only one instance of uncritical processes, thus having fewer processes in total.
In addition, the joint use of larger shared memory buffers yields synergy effects, such that more time frames can be in flight with the same memory constraint.

Currently, the asynchronous reconstruction runs only the TPC tracking on the GPU, since the latest version of the ITS tracking algorithm is not compatible with GPUs.
The ITS tracking was improved in several aspects, and GPU compatibility was broken to simplify the development.
GPU compatibility will be restored in the near future.
From Table~\ref{tab:asyncreco}, around 60\% of the compute load from the TPC tracking can currently run on the GPUs, reducing the CPU compute time by 60\%, allowing for a speedup factor of 2.5.
Comparing the best GPU against the best CPU configuration in Table~\ref{tab:asyncperf}, this factor of 2.5 is almost exactly reached with 4.27s against 1.7s.
Thus, under the given constraints, the usage of GPUs in the asynchronous reconstruction works optimally.

The aggregate compute load of the barrel tracking in Table~\ref{tab:asyncreco} is around 80\% of the total asynchronous compute load.
Thus, once the full optimistic scenario is implemented on the GPU, the speedup should increase to a factor 5.
The fraction of compute power (measured using TPC tracking) of the EPNs that is provided by the GPUs is 85\% of the total EPN compute power, thus this speedup factor of 5 is realistically achievable.
Thinking beyond the optimistic scenario, it would not make sense to port more than 85\% of the workload to the GPU, since then one would become GPU-bound.
The absolute limit for the GPU speedup is a factor of around 6.5, if exactly 85\% of the workload would run on the GPU.

\section{Conclusions}

For Run 3, ALICE has switched to a new common software framework DPL for synchronous and asynchronous processing of the data, instead of the former separate online and offline frameworks.
The synchronous processing relies heavily on GPUs to provide the compute power necessary for processing heavy-ion collision data in real time at the highest rates.
Today, around 99\% of the compute load of the synchronous processing is running on the GPUs.
The online processing farm consists of 350 servers with 8 GPUs each, and according to Monte Carlo simulations, it can handle 50 kHz Pb--Pb data with a compute margin of 30\%.
A comparable CPU-only farm would require more than 2000 servers with 64 cores each.
At the moment, 60\% of the compute load of asynchronous processing is ported to GPUs, yielding a speedup factor of 2.5 on EPN servers compared to the CPU-only workflow.
To optimise the EPN server load during asynchronous processing, ALICE aims to port the full barrel tracking chain to GPUs, which will offload 80\% of the compute load and should yield a speedup factor of 5.
During pp data taking in 2021 to 2023, and during the low interaction rate heavy-ion run in 2022, the data processing performed as expected.
The first operation at the designed peak load will happen in the end of 2023, when ALICE will process Pb--Pb collisions at 50 kHz interaction rate for the first time.

\bibliography{bibliography}

\end{document}